\documentclass{ws-procs9x6}
\def\fsl#1{\setbox0=\hbox{$#1$}                 
   \dimen0=\wd0                                 
   \setbox1=\hbox{/} \dimen1=\wd1               
   \ifdim\dimen0>\dimen1                        
      \rlap{\hbox to \dimen0{\hfil/\hfil}}      
      #1                                        
   \else                                        
      \rlap{\hbox to \dimen1{\hfil$#1$\hfil}}   
      /                                         
      \fi}                                      %

\newcommand{\VEV}[1]{\langle #1 \rangle}

\begin{document}

\title{Scalar Decay Constant and Yukawa Coupling \\ in Walking
  Technicolor Models}
\author{Michio Hashimoto$^*$}
\address{Chubu University, \\
1200 Matsumoto-cho, Kasugai-shi, \\
Aichi, 487-8501, JAPAN \\
$^*$E-mail: michioh@isc.chubu.ac.jp}

\begin{abstract}
Based on Refs.~\refcite{Hashimoto:2011ma} and \refcite{Hashimoto:2011cw},
we study the couplings of the scalar bound state
to the fermions and the weak bosons in walking gauge theories.
\end{abstract}


\bodymatter

\section{Introduction}\label{sec1}

Recently, a modest excess of events around the Higgs mass, 
$m_h \sim 125$~GeV, over the standard model (SM) background
has been reported~\cite{LHC-Higgs-search}.
This Higgs mass is consistent with the precision measurements~\cite{pdg}.
I would like to mention, however, it is not yet conclusive.
The mechanism for the electroweak symmetry breaking is still unrevealed.

Based on Refs.~\refcite{Hashimoto:2011ma} and \refcite{Hashimoto:2011cw},
we study the couplings of the scalar bound state,
so-called the technidilaton (TD), to the SM fermions and the weak bosons 
in walking technicolor (WTC). 
These are crucial for the TD searches.

\section{Coupling to the SM fermions}

\begin{figure}[t]
  \begin{center}
  \resizebox{0.15\textwidth}{!}{\includegraphics{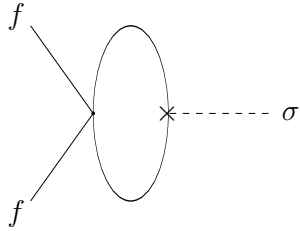}}
  \end{center}
  \caption{Yukawa coupling between the SM fermions $f$ and 
   the scalar bound state $\sigma$ in ETC. 
   The TF loop generates the mass of $f$ and 
   also intermediates between $f$ and $\sigma$.
  \label{fig-yukawa}}
\end{figure}

Suppose that the extended technicolor (ETC) sector generates
the four-fermion interaction and that 
the SM fermion mass $m_f$ is obtained from the technifermion (TF) condensate,
$\VEV{\bar{\psi}\psi}$.
See also Fig.~\ref{fig-yukawa}.
By introducing the scalar decay constant $F_S$ for the scalar current, 
$ \langle 0|(\bar{\psi}\psi(0))_R| \sigma (q) \rangle \equiv 
 F_S M_\sigma $, 
where $M_\sigma$ is the mass of the scalar bound state $\sigma$
and the subscript $R$ represents the renormalized quantity, 
we can then obtain the yukawa coupling~\cite{Hashimoto:2011ma},
\begin{equation}
  g_{\sigma ff} = 
  \frac{m_f}{\frac{-\VEV{\bar{\psi}\psi}_R}{F_S M_\sigma}} \, .
  \label{yukawa-Ftd}
\end{equation}

We perform the calculations of $F_S$ and $\VEV{\bar{\psi}\psi}_R$
by using the improved ladder SD equation~\cite{Hashimoto:2010nw}. 
We then obtain
\begin{equation}
  \frac{g_{\sigma ff}}{g_{hff}^{\rm SM}} = \sqrt{\frac{N_{\rm TC}}{N_f}} N_D 
  \frac{\kappa_F^2 \sqrt{5-\tilde{\omega}^2}}{4\pi\sqrt{2\kappa_V^{}}}
  \frac{M_\sigma}{v} , 
\end{equation}
where the SM yukawa coupling is 
$g_{hff}^{\rm SM} = m_f/v$ with $v = 246$~GeV. 
Also, $N_{\rm TC}$, $N_f$ and $N_D$ denote the number of the color
of the TC gauge group, the number of the flavor and the weak doublets 
for each TC index, respectively.
The values of $\kappa_F$ and $\kappa_V$ are defined by
\begin{equation}
  v^2 = N_D F_\pi ^2 \equiv \frac{N_{\rm TC} N_D}{4\pi^2} \, \kappa_F^2\, m^2 ,
 \quad \mbox{and} \quad
   \VEV{\theta_\mu^\mu} \equiv 
  -\frac{N_{\rm TC} N_f}{2\pi^2} \kappa_V^{} m^4 , 
  \label{k_F}
\end{equation}
where $m$ is the dynamically generated TF mass. 
We show the numerical values of 
$g_{\sigma ff}/g_{hff}^{\rm SM}$ in Table~1. 
We here used the WTC relation $N_f \simeq 4N_{\rm TC}$ and
$\Lambda_{\rm ETC}$ represents the ETC scale.
The values of $\tilde{\omega}$ are obtained through those of $\lambda_*$,
$\tilde{\omega} \equiv \sqrt{4\lambda_*-1}$.
\begin{equation*}
  \begin{array}{c|c|c|c||c||c}\hline
    \lambda_* & \frac{m}{\Lambda_{\rm ETC}} & \kappa_V^{} & \kappa_F^{} &
     \sqrt{\frac{N_D}{N_f}} \frac{F_S}{v} & 
     \frac{g_{\sigma ff}}{g_{h ff}^{\rm SM}} \frac{v}{N_D M_\sigma} \\ \hline
   0.305 & 1.12 \times 10^{-3} & 0.685 & 1.38 & 2.59 & 0.142 \\
   0.287 & 1.08 \times 10^{-4} & 0.709 & 1.42 & 2.71 & 0.148 \\
   0.258 & 5.88 \times 10^{-10} & 0.756 & 1.48 & 2.93 & 0.157 \\ \hline
 \end{array}
\end{equation*}
\centerline{Table 1. ~~Numerical values of $g_{\sigma ff}/g_{hff}^{\rm SM}$ .}

For the typical one-family TC model with $N_{\rm TC}=2$, $N_f=8$
and $N_D=4$, we can read $m=$ 390~GeV, 380~GeV, 370~GeV from top 
to bottom in Table~1.
The handy Higgs mass formula~\cite{Hashimoto:1998qk}, 
$M_\sigma \simeq \sqrt{2}m$,
then yields $M_\sigma \simeq$ 560~GeV, 540~GeV, 520~GeV, respectively.
For the typical Higgs mass, $M_\sigma = $ 500~GeV,
we obtain $g_{\sigma ff}/g_{hff}^{\rm SM} \simeq 1.2$.
Furthermore, there are $2N_{\rm TC} \; (=4)$ extra colored fermions
(techniquarks).
Therefore the production cross section of $\sigma$ in such a model should be
considerably enhanced, 
like in the fourth generation models~\cite{He:2001tp}.
It has been severely constrained by 
the recent LHC data~\cite{LHC-Higgs-search}.
On the other hand, it is not the case 
for the model having only one weak doublet and no extra techniquark. 
We also note that signatures of some classes of the top condensate 
models~\cite{Hill:2002ap} are similar to the SM.

\section{Coupling to the weak bosons}

We may regard the scalar bound state $\sigma$ as a dilaton.

When the dilaton $\sigma$ directly couples to $W$,
like in the SM, one can easily derive the $\sigma$--$W$--$W$ 
coupling as~\cite{Goldberger:2007zk} 
\begin{equation}
   g_{\sigma WW} = \frac{2M_W^2}{F_\sigma}, \label{sWW}
\end{equation}
where $F_\sigma$ represents the dilaton decay constant being
$\langle 0| \theta_\lambda^\lambda (0)|\sigma (q) \rangle = F_\sigma M_\sigma^2$.
Notice that $F_S$ in the previous section is different from $F_\sigma$.

Next, we consider the situation that 
the TD couples to the weak bosons only through the TF loop.

\begin{figure}[t]
   \begin{center}
   \resizebox{0.57\textwidth}{!}{\includegraphics{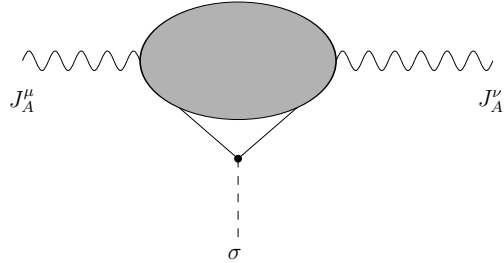}}
   \end{center}
   \caption{Coupling of the TD to the axial current $J_A^\mu$ of the TF.
   The TD couples to $J_A^\mu$ only through the internal TF lines.
   \label{fig-sAA}}
\end{figure}

Since the axial current $J_A^\mu$ of the TF's yields
the decay constant $F_\pi$, 
$\langle 0| J_A^\mu(0)|\pi (q) \rangle = -iq^\mu F_\pi$,
and the weak boson mass is provided by $F_\pi$, 
the coupling between $\sigma$ and $J_A^\mu$ should be crucial.
See also Fig.~\ref{fig-sAA}.

The axial current correlator in the momentum space is 
\begin{equation}
  {\rm F.T.}i\langle 0| J_A^\mu(x) J_A^\nu(0)|0\rangle = 
  \left(g^{\mu\nu}-\frac{q^\mu q^\nu}{q^2}\right)\Pi_A(q^2) , \quad
  \mbox{and} \quad \Pi_A (0) = F_\pi^2 , 
  \label{Pi_A}
\end{equation}
which plays an important role in our approach.

The $\sigma$ coupling to $J_A^\mu$ at the zero momentum transfer
is just like the mass insertion:
Note that the identity holds
\begin{equation}
  \frac{1}{\fsl{\ell} - m} y_T^{} \frac{1}{\fsl{\ell} - m} 
  = y_T^{} \frac{\partial}{\partial m} \frac{1}{\fsl{\ell} - m} ,
\end{equation}
where $y_T^{}$ represents the yukawa coupling between the TD and the TF.
We can then obtain the coupling of $\sigma$ to $J_A^\mu$
at zero momentum simply by
\begin{equation}
  g_{\sigma AA}(0) = y_T^{} \frac{\partial \Pi_A(0)}{\partial m}\,.
  \label{sT-AA} 
\end{equation}
Because $F_\pi$ is expected to be proportional to $m$, i.e., 
$F_\pi = \kappa \, m$, with
$\kappa \equiv \kappa_F \sqrt{N_{\rm TC}}/(2\pi)$ and
$\kappa_F \simeq 1.4\mbox{--}1.5$ in Eq.~(\ref{k_F}),
Eqs.~(\ref{Pi_A}) and (\ref{sT-AA}) then yield 
$g_{\sigma AA}(0) = y_T^{} \cdot 2F_\pi^2/m$.
Attaching $W^\mu$ to $J_A^{\mu}$,
we finally obtain the coupling of the TD to the weak bosons
at zero momentum,
\begin{equation}
  g_{\sigma WW}(0) = y_T^{} \frac{2M_W^2}{m} \, .
  \label{sT-WW}
\end{equation}

When the yukawa coupling is like the SM, $y_T^{} = m/F_\sigma$,
Eq.~(\ref{sT-WW}) formally agrees with Eq.~(\ref{sWW}).
For the model in Ref.~\refcite{Bando:1986bg},
where the four-fermion interactions were incorporated,
$y_T$ was estimated as $y_T^{} = (3-\gamma_m)m/F_\sigma$
with $\gamma_m \simeq 1$.
If so, $g_{\sigma WW}$ is changed by 
the additional factor $(3-\gamma_m)$. 
In any case, we conclude that 
the (effectively induced) operator
$\frac{\sigma}{F_\sigma} W_\mu W^\mu$ yields 
the coupling between the TD and the weak bosons,
similarly to the SM.

\section{Summary}

We studied the couplings of $\sigma$ to $f$ and $W$.
For details, see Refs.~\refcite{Hashimoto:2011ma} and 
\refcite{Hashimoto:2011cw}.

\end{document}